\documentclass[proof]{WileyASNA-v1}

\usepackage{tikz, graphicx}	% Including figure files
\usepackage{amsmath}	% Advanced maths commands
\usepackage{hyperref}
\usepackage{tabularx}
\usepackage{subcaption}
\usepackage{wasysym}
\usepackage[
    starfontserif % comment for sans glyphs
    ]{starfont}

\DeclareSymbolFont{starfontsym}{OT1}{sts}{m}{n}
\DeclareMathSymbol{\mathSun}{\mathord}{starfontsym}{115}
\DeclareMathSymbol{\mathMercury}{\mathord}{starfontsym}{102}
\DeclareMathSymbol{\mathVenus}{\mathord}{starfontsym}{103}
\DeclareMathSymbol{\mathTerra}{\mathord}{starfontsym}{76}
\DeclareMathSymbol{\mathvarTerra}{\mathord}{starfontsym}{108}
\DeclareMathSymbol{\mathMoon}{\mathord}{starfontsym}{100}
\DeclareMathSymbol{\mathvarMoon}{\mathord}{starfontsym}{97}
\DeclareMathSymbol{\mathMars}{\mathord}{starfontsym}{104}
\DeclareMathSymbol{\mathJupiter}{\mathord}{starfontsym}{106}
\DeclareMathSymbol{\mathSaturn}{\mathord}{starfontsym}{83}
\DeclareMathSymbol{\mathUranus}{\mathord}{starfontsym}{70}
\DeclareMathSymbol{\mathvarUranus}{\mathord}{starfontsym}{65}
\DeclareMathSymbol{\mathNeptune}{\mathord}{starfontsym}{71}
\DeclareMathSymbol{\mathPluto}{\mathord}{starfontsym}{74}
\DeclareMathSymbol{\mathvarPluto}{\mathord}{starfontsym}{72}

\defcitealias{2024AN....34540039B}{Paper\,I}

\graphicspath{{images/}}
\articletype{Original Article}

\received{19 Nov 2024}
\revised{14 Jan 2025}
\accepted{15 Jan 2025}

\begin{document}

\title{
\texorpdfstring{
\textit{JWST} Imaging of the Closest Globular Clusters -- V.\\ 
The White Dwarfs Cooling Sequence of M\,4\protect\thanks{
Based on observations with the  NASA/ESA {\it James Webb Space Telescope},
obtained at the  Space Telescope Science Institute,  which is operated
by AURA, Inc., under NASA contract NAS 5-26555, under GO-1979.
}}
{JWST on NGC 6397 - White Dwarfs}
}

% R: so far had worked to the manuscript
\author[1]{L. R. Bedin}
\author[1]{M. Libralato}
\author[2]{M. Salaris}
% R: team-members that worked more or less indirectly to the software and calibration used in the paper
\author[1,3]{D. Nardiello}  % PSFs
\author[1,4]{M. Scalco}     % Testing 
\author[1,5]{M. Griggio}    % GDCs 
\author[5]{J. Anderson}     % core-build of the software
%
% R: other team-members and Co-IS
\author[6]{P. Bergeron}
\author[5]{A. Bellini}
\author[7]{R. Gerasimov}
\author[8]{A. J. Burgasser}
\author[9]{D. Apai} 

\authormark{\textsc{L. R. Bedin et al.}}

\address[1]{\orgdiv{Istituto Nazionale di Astrofisica}, \orgname{Osservatorio Astronomico di Padova}, \orgaddress{\state{Vicolo dell'Osservatorio 5, Padova, IT-35122}, \country{Italy}}}
\address[2]{Astrophysics Research Institute, Liverpool John Moores University, 146 Brownlow Hill, Liverpool L3 5RF, UK}
\address[3]{\orgdiv{Universit\`a di Padova}, \orgname{Dipartimento di Fisisca e Astronomia}, \orgaddress{\state{Vicolo dell'Osservatorio 3, Padova, IT-35122}, \country{Italy}}}
\address[4]{\orgdiv{Universit\`a di Ferrara}, \orgname{Dipartimento di Fisica}, \orgaddress{\state{Via Giuseppe Saragat 1, I-44122, Ferrara}, \country{Italy}}}
\address[5]{\orgname{Space Telescope Science Institute}, \orgaddress{\state{3800 San Martin Drive, Baltimore,  MD 21218}, \country{USA}}}
\address[6]{\orgname{D\'epartement de Physique, Université de Montr\'eal}, \orgaddress{\state{C.P. 6128, Succ. Centre-Ville, Montr\'eal,  QC H3C 3J7}, \country{Canada}}}
\address[7]{\orgdiv{Department of Physics and Astronomy}, \orgname{University of Notre Dame}, \orgaddress{\state{Notre Dame, Nieuwland, Science Hall, IN, 46556}, \country{USA}}}
\address[8]{\orgdiv{Department of Astronomy \& Astrophysics}, \orgname{University of California San Diego}, \orgaddress{\state{La Jolla, CA 92093}, \country{USA}}}
\address[9]{\orgdiv{Department of Astronomy and Steward Observatory}, \orgname{The University of Arizona}, \orgaddress{\state{933 N. Cherry Avenue, Tucson, AZ 85721}, \country{USA}}}

\corres{*E-mails: luigi.bedin@inaf.it}

%%%\presentaddress{This is sample for present address text this is sample for present address text}

%%%%%%%%%%%%%%%%%%%%%%%
\abstract{
We combine infrared (IR) observations collected by the \textit{James Webb Space Telescope}  
with optical deep images by the \textit{Hubble Space Telescope} taken 
approximately 20\,years earlier to compute proper-motion 
membership for the globular cluster (GC) M\,4 (NGC\,6121) 
along its entire white dwarf (WD) cooling sequence (CS). 
These new IR observations allow us,  for only the second time in a GC, 
to compare WD models with observations over a wide range of wavelengths,  
constraining fundamental astrophysical properties of WDs.
Furthermore, we investigate the presence of WDs with IR excess along the WD\,CS of M\,4,    
similar to the recent study conducted on the GC NGC\,6397.
We also determine  the age difference between M\,4 and NGC\,6397 
by comparing the absolute F150W2 magnitudes of the luminosity function peak 
at the bottom of the observed  WD\,CS, and find that M\,4 is slightly younger, by 0.8$\pm$0.5\,Gyr.} 
%}  
%
%}
%
%%%%%%%%%%%%%%%%%%%%%%%

\keywords{globular cluster (individual): M\,4 (NGC\,6121), astrometry, photometry: white dwarfs}

\fundingInfo{
This work is based on funding by: 
INAF under the WFAP project, f.o.:1.05.23.05.05;  
The Science and Technology Facilities Council Consolidated Grant ST/V00087X/1; and 
STScI funding associated with GO-1979. 
}

\maketitle

%%%fig
\footnotetext{\textbf{Abbreviations:} 
\textit{JWST}, James Webb Space Telescope; 
\textit{HST}, Hubble Space Telescope; 
%$\mathJupiter$, Jupiter; 
%$\mathNeptune$, Neptune; 
%$\mathTerra$, Earth;
$\mathSun$, Sun/Solar
}
%%%
%%%%%%%%%%%%%%%%%%%%%%%%%%%%%%%%%%%%%%%%%%%%%%%%%%%%%%%%%%%%%%%%%%%%%%%%%%%%%%%
%%%%%%%%%%%%%%%%%%%%%%%%%%%%%%%%%%%%%%%%%%%%%%%%%%%%%%%%%%%%%%%%%%%%%%%%%%%%%%%
%%%%%%%%%%%%%%%%%%%%%%%%%%%%%%%%%%%%%%%%%%%%%%%%%%%%%%%%%%%%%%%%%%%%%%%%%%%%%%%
\section{Introduction}\label{S:intro}
%
% %
%%%%%%%%%%%%%
Globular clusters (GCs), among the oldest objects in the Universe, 
serve as ideal laboratories for 
testing stellar evolution models, derive crucial information about
galaxy formation and cosmology \citep[see, e.g., the reviews by][]{kr, forbes}, thanks
to their stars' uniform age, distance, and chemical 
composition \citep[at least to a first approximation;][]{2015AJ....149...91P}. 
In this context, the colour-magnitude diagrams (CMDs) 
of their stellar components are one of the most crucial tools 
for such investigations. 
To render CMDs truly informative, proper motions (PMs) are fundamental 
to establishing cluster membership, particularly for their faintest stars, 
including the white dwarf (WD) population.\\ 

Without PM measurements, these stars would be 
otherwise irremediably lost and confused with the multitude of 
foreground and background field objects surrounding the GCs.

By leveraging the astrometric capabilities of the 
\textit{James Webb Space Telescope (JWST)} \citep{2023AN....34430006G} 
alongside deep \textit{Hubble Space Telescope (HST)} images 
collected approximately 20 years ago, we can now 
explore the stellar components of GCs over a wide photometric spectral range, and 
more importantly, obtain precise PM measurements.

Our \textit{JWST} program GO-1979 \citep{2021jwst.prop.1979B}, 
was specifically designed to measure high-precision infrared photometry and 
astrometry of the faintest objects in the two nearest Galactic GCs, namely  
% https://people.smp.uq.edu.au/HolgerBaumgardt/globular/fits/ngc6121.html
% https://people.smp.uq.edu.au/HolgerBaumgardt/globular/fits/ngc6397.html
Messier\,4 (M\,4, a.k.a. NGC\,6121, at a distance $d$=1.85$\pm$0.02\,kpc) and NGC\,6397 \citep[$d$=2.48$\pm$0.02\,kpc;][]{2021MNRAS.505.5957B}.
Due to their proximity and different 
metallicities ([Fe/H]=$-$1.18 for M\,4  and [Fe/H]=$-$1.99 for NGC\,6397, see \citealt{carretta09}) 
they are the first natural targets 
to investigate the infrared properties of the WD cooling sequences 
of Galactic GCs, and if/how they change with the cluster initial chemical composition.

A series of papers \citep[presented and summarized by ][]{2024AN....34540039B} 
studied different aspects of NGC\,6397 based on our \textit{JWST} GO-1979 program. \citep[Paper\,I;][]{2024AN....34540039B} has studied the white dwarfs (WDs), while the 
brown dwarf population was presented in \citet[Paper\,II;][]{2024ApJ...971...65G}. The multiple population phenomenon was studied in \citet[Paper\,III;][]{2024A&A...689A..59S}. Our fourth paper \citep[Paper\,IV;][]{2024arXiv240906774L} investigated the faint main sequence (MS) chemistry in the parallel fields of both clusters. 
This fifth paper of the series is focused on the 
WDs of M\,4.

The WD cooling sequence (CS) of M\,4 
was previously studied with deep \textit{HST} optical 
photometry by \citet{hansen02}, \citet{hans04M4}, and 
down to its bottom by \citet{2009ApJ...697..965B}.
These works were all centred on the age determination of the cluster from its WD cooling sequence. Such studies take advantage of the fact that WDs fade with time towards progressively fainter magnitudes. The older the WD the dimmer its
magnitude, hence the CMD of the WD population in a star cluster is
expected to display a cut-off at a certain magnitude. This cut-off magnitude is a function of the cluster age. \citet{2009ApJ...697..965B} determined an age of 
11.6$\pm$0.6~Gyr (internal errors only) whilst 
\citet{hans04M4} -- improving upon the earlier analysis 
by \citet{hansen02} with the same data, taking into account 
various possible sources of systematic uncertainties -- obtained a best estimate of 12.1~Gyr, with a 95\% lower 
limit of 10.3~Gyr.
%
%%%%%%%%%%%%%%%%%%%%%%%%%%%%%%%%%%%%%%%%%%%%%%%%%%%%%%%%%%%%%%%%%%%%%%%%%%%%%%%
%%%%%%%%%%%%%%%%%%%%%%%%%%%%%%%%%%%%%%%%%%%%%%%%%%%%%%%%%%%%%%%%%%%%%%%%%%%%%%%
%%%%%%%%%%%%%%%%%%%%%%%%%%%%%%%%%%%%%%%%%%%%%%%%%%%%%%%%%%%%%%%%%%%%%%%%%%%%%%%
%__________________________________________________________________
\begin{figure*}[t]
\centerline{
\includegraphics[height=54mm]{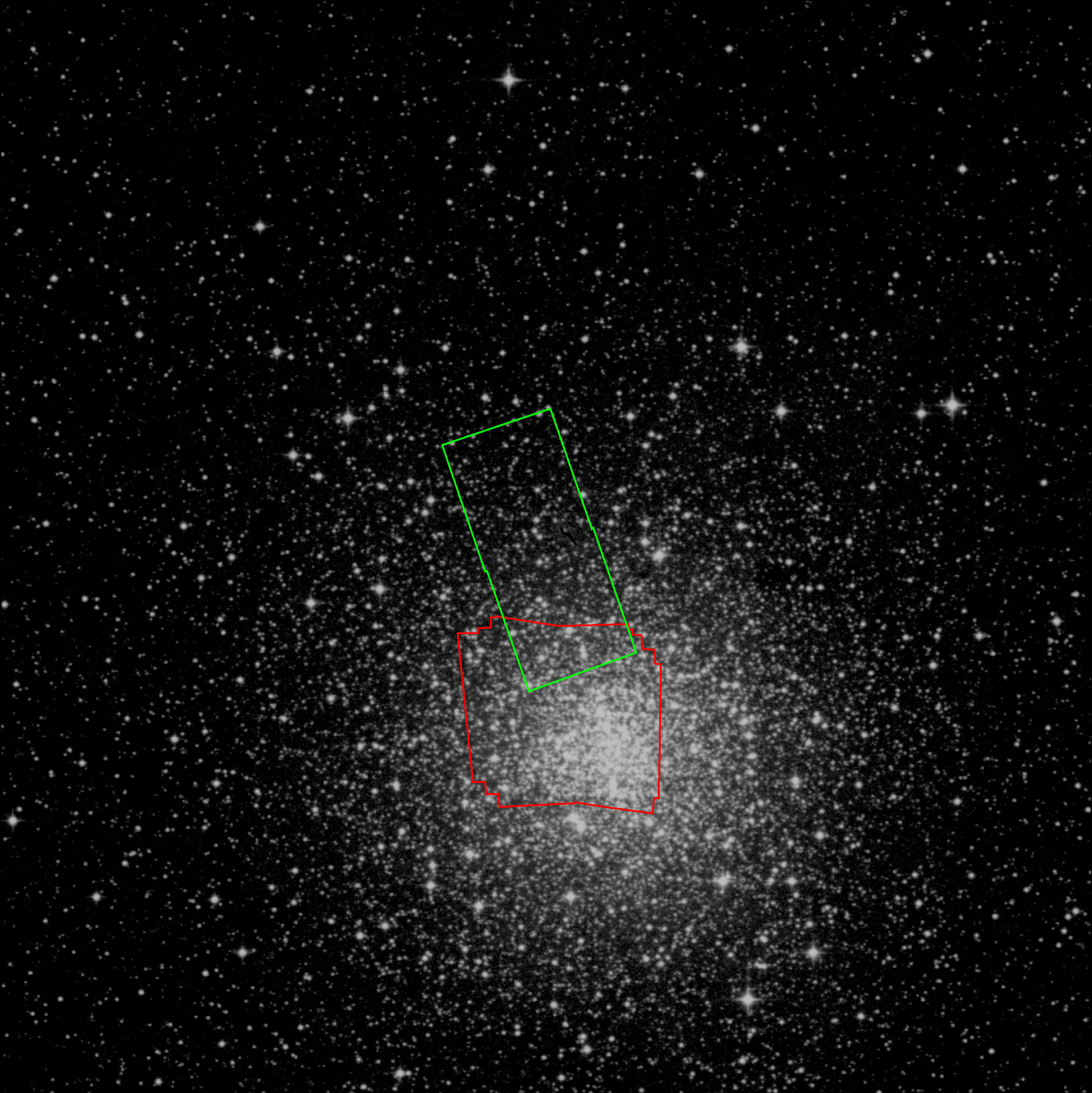}
\includegraphics[height=54mm]{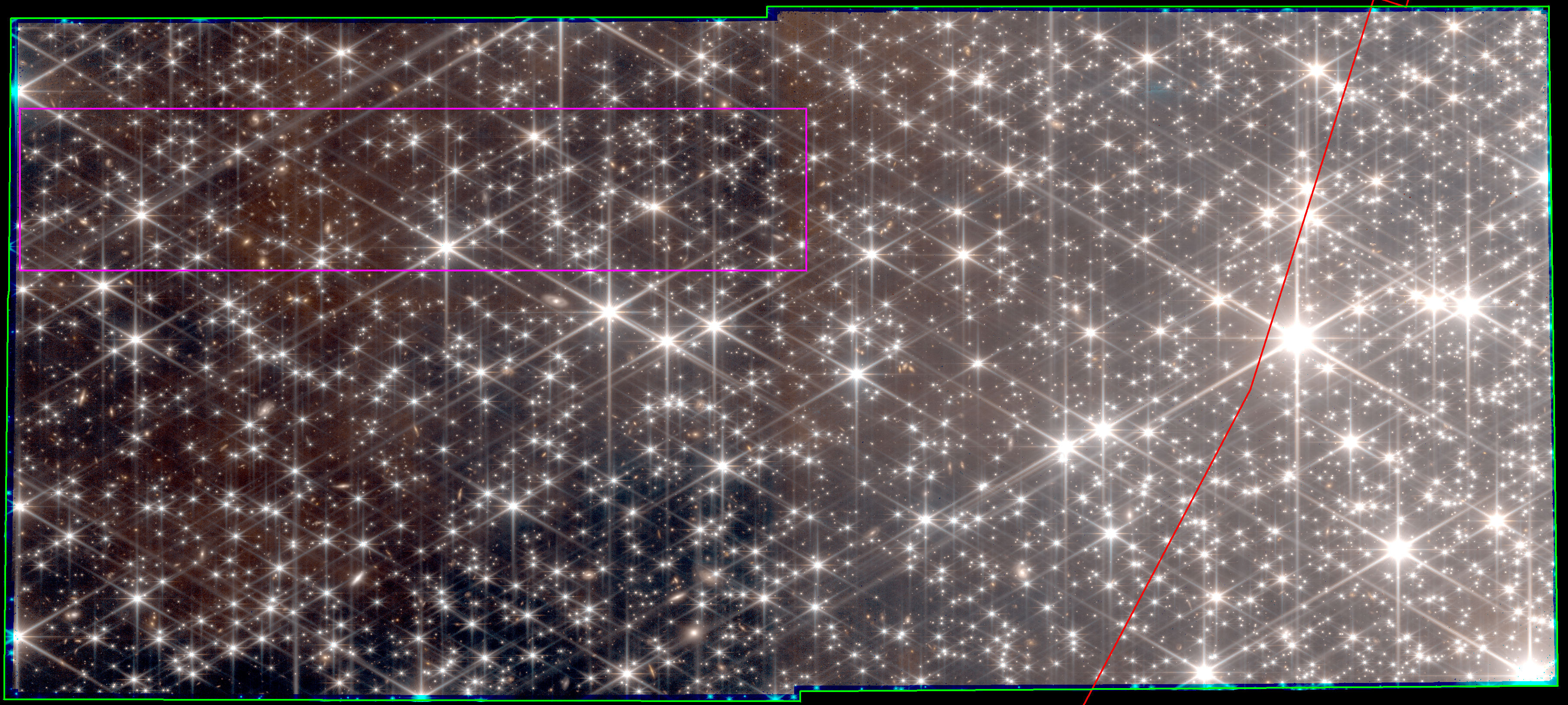}
}
~\\
\centerline{
\includegraphics[width=176mm]{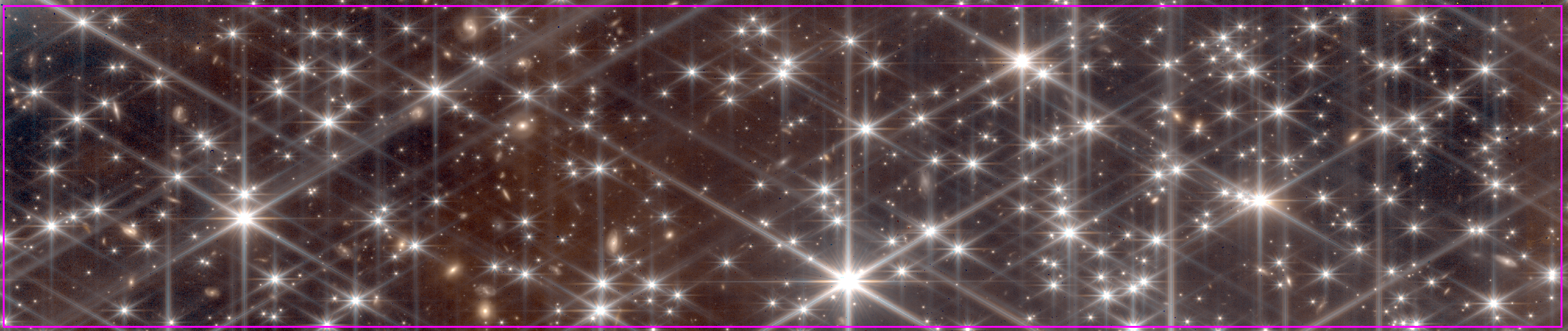}
}
\caption{
%%%
\textit{(Top-Left:)} 
A $ 25^\prime\times 25^\prime $ 
infrared image from the Digital\,Sky\,Survey\,2 centered on our NIRCam field of M\,4 for \textit{JWST} 
program GO-1979 (green box). The image is aligned with North up and East toward the left.
The region indicated in red shows the archival \textit{HST} deep field from programs GO-10146. 
\textit{(Top-Right:)} The entire NIRCam primary field.
\textit{(Bottom:)} A zoom-in on a representative dark sub-region (of $\sim 3^\prime\times 0.6^\prime$ ) in 
the NIRCam image (highlighted in magenta in the top panel). 
}
%%%
\label{fig:FoV}
\end{figure*}
%__________________________________________________________________
%
%

In this paper, we investigate, for only the second time in the infrared (IR) within a GC ---following NGC 6397--- the effects of 
collision-induced absorption (CIA) of H$_2$ molecules in the cool atmospheres of WDs.  
Unlike in the optical, these effects are unambiguous in this wavelength 
range \citepalias[see][]{2024AN....34540039B}.

Furthermore, we investigate whether M\,4 
shows evidence of a WD subpopulation 
with excess flux in the F322W2 filter (up to $\sim$0.5\,magnitudes), 
previously identified in NGC\,6397 WDs \citepalias[][]{2024AN....34540039B}.

The excess observed among NGC\,6397's WDs requires further confirmation, as it is limited to the F322W2 filter 
and becomes more pronounced at fainter magnitudes. 
In \citetalias{2024AN....34540039B} several potential explanations for this IR excess were explored, including a 
large population of WD+BD binaries, reddening from circumstellar debris disks or post-AGB material, 
contributions from helium-dominated or helium-core WDs, and a possible link to the two 
populations in NGC\,6397's stellar main sequence; however, 
no definitive explanation was identified.
Building on our results for NGC\,6397, we now aim to investigate whether a similar infrared flux excess 
is present in M\,4. 
This investigation could be crucial, as expanding our sample of GCs will enhance 
our understanding of the mechanisms behind 
these intriguing observations.

In addition, we revisit the cluster age derived from the cluster CS, by comparing its cut-off magnitude with that of the metal poorer cluster NGC\,6397 investigated in detail in Paper\,I. The age difference between these two clusters 
has important implications for the age-metallicity relation of the Galactic GC system, which is related to the formation of the Galactic halo \citep[see, e.g.,][]{leaman13}.

Age determinations based on the main sequence turn-off 
brightness provided so far inconsistent results, with 
M\,4  younger by an amount ranging  
from less than 1~Gyr \citep[e.g.,][]{sw, mf} to 
1.5~Gyr \citep{vdb}.
The study of NGC\,6397 CS by \citet{hansen07} provides 
an age equal to 11.5$\pm$0.5~Gyr (95\% confidence limits), that is consistent, within the errors (that are larger for M\,4), with M\,4 age derived from its CS by 
\citet{hans04M4}.

Here we employ the most recent determinations of the clusters' distances \citep{2021MNRAS.505.5957B}, the power of infrared photometry to minimize the effect of extinction, which is high for these two clusters (
$E(B-V)\sim$0.2 for NGC~6397 and $\sim$0.4 for M\,4, whose extinction is also characterized by a ratio $R_V=A_V/E(B-V)$ larger than the standard value of 3.1, 
see, e.g., 
\citealt{2012AJ....144...25H}), and the updated WD \citep{bastiwd} and progenitor models 
\citep{bastiiacaen} from the BaSTI-IAC database.\\  

The paper is organized as follows: 
Section\,2 presents the observations; 
Section\,3 describes the data reduction, proper-motion analysis, and artificial star tests; 
Section\,4 discusses the observational results on the WD CS of M 4 and their interpretation. 
Finally, our conclusions are outlined in Section\,5. 
As part of this publication, we also publicly release supplementary online materials,
including both the astrometrized atlases of the studied field and the photometric catalogues.

%%%%%%%%%%%%%%%%%%%%%%%%%%%%%%%%%%%%%%%%%%%%%%%%%%%%%%%%%%%%%%%%%%%%%%%%%%%%%%%
%%%%%%%%%%%%%%%%%%%%%%%%%%%%%%%%%%%%%%%%%%%%%%%%%%%%%%%%%%%%%%%%%%%%%%%%%%%%%%%
%%%%%%%%%%%%%%%%%%%%%%%%%%%%%%%%%%%%%%%%%%%%%%%%%%%%%%%%%%%%%%%%%%%%%%%%%%%%%%%
%__________________________________________________________________
\begin{figure*}[t]
\centerline{
\includegraphics[height=53mm]{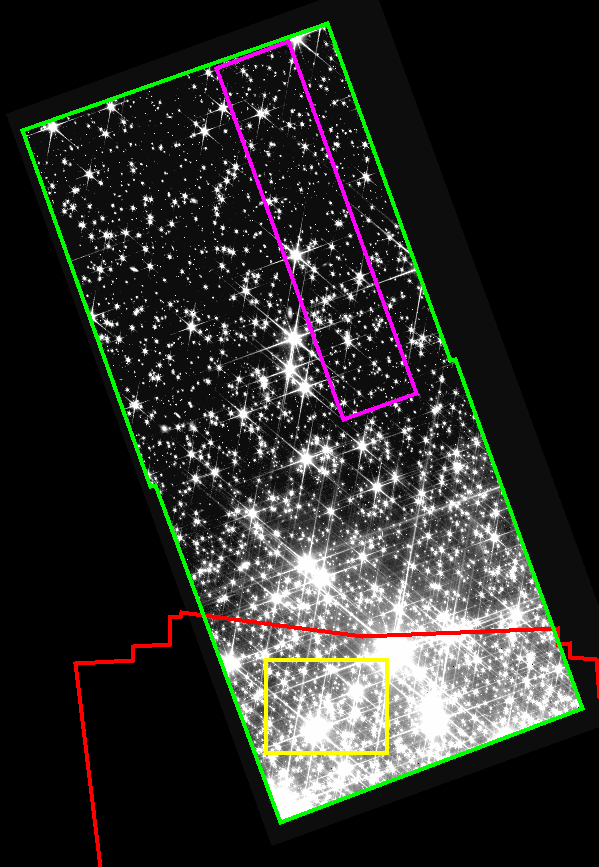}
\includegraphics[height=53mm]{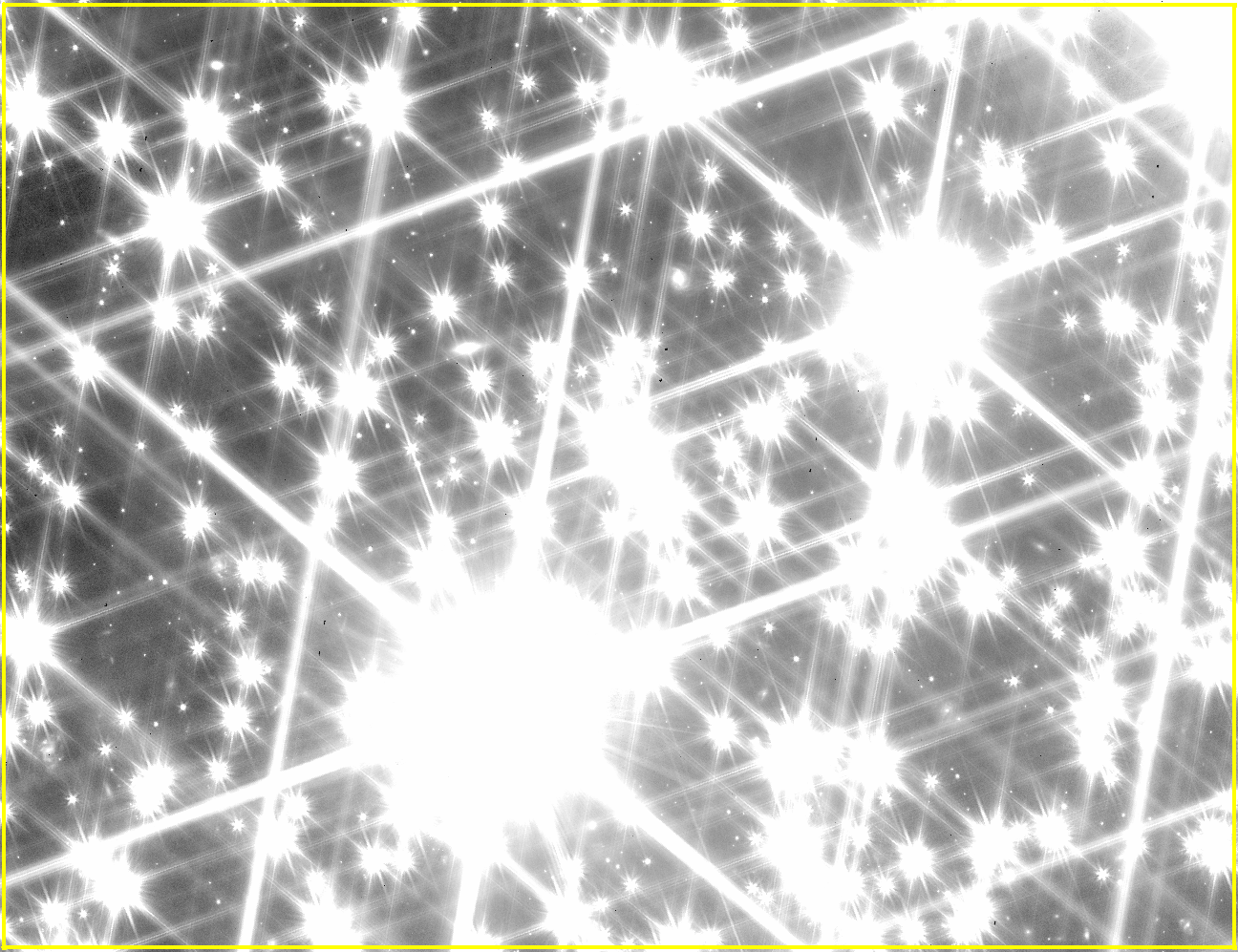}
\includegraphics[height=53mm]{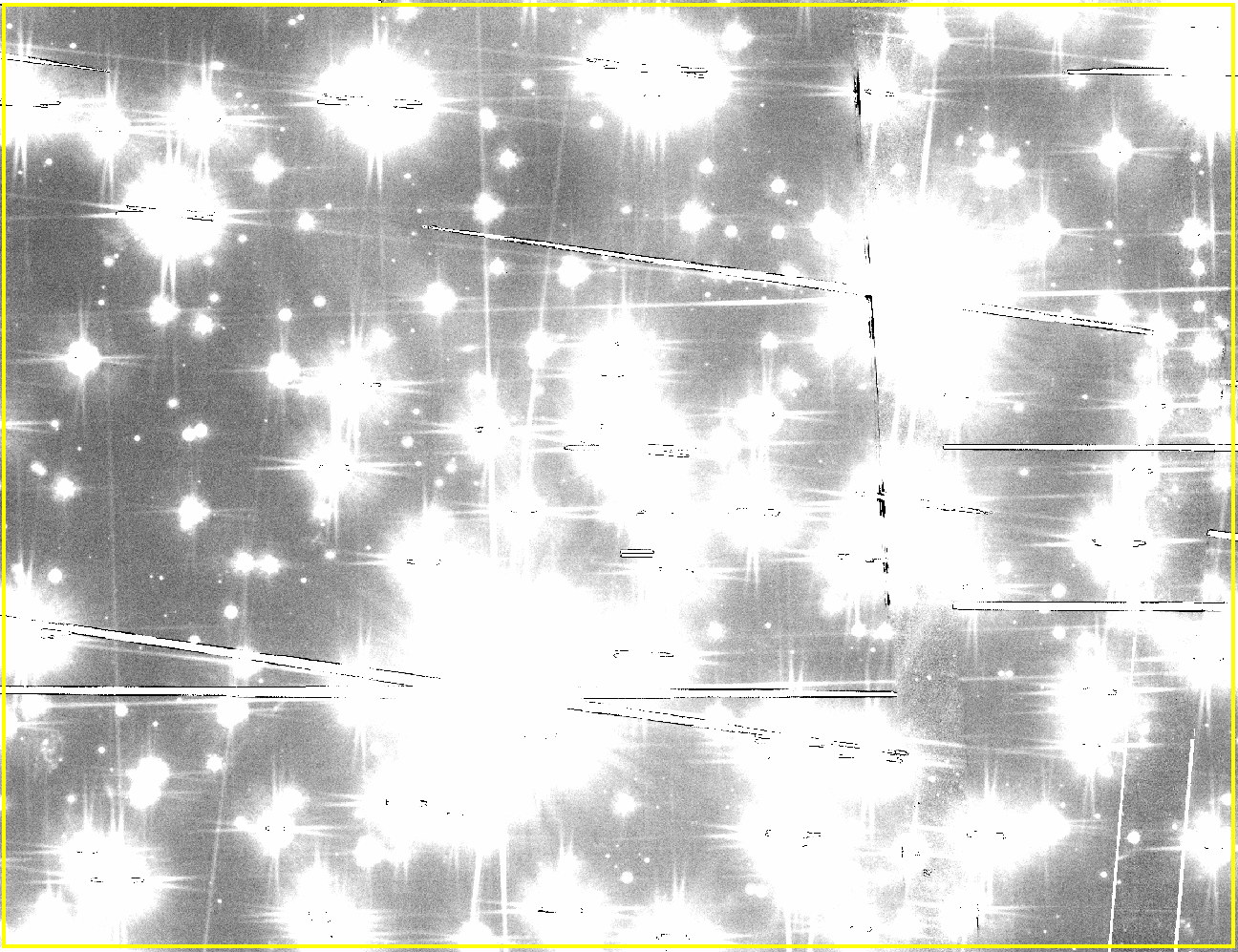}
}
\caption{
%%%
\textit{(Left-Panel:)} This figure highlights the difference in the amount of scattered light 
between ACS/WFC@HST and NIRCam/SW images. We define a common region between the two datasets and mark it in yellow.
\textit{(Mid-Panel:)} Zoom-in of the region indicated in yellow as seen with \textit{JWST}/NIRCam/SW/F150W2. 
\textit{(Right-Panel:)} Same region, as seen with \textit{HST}/ACS/WFC/F775W.
In these panels, the scale is stretched to reveal the fluctuations of the sky background in the 
darkest areas not dominated by the halos of bright stars. 
Clearly, \textit{JWST} images have less regions than \textit{HST} free from spikes and halos 
from the brightest stars, which makes it difficult to detect the faintest sources which 
should compete not only against the sky brightness but also against other features.  
}
%%%
\label{fig:Zom}
\end{figure*}
%__________________________________________________________________

%%%%%%%%%%%%%%%%%%%%%%%%%%%%%%%%%%%%%%%%%%%%%%%%%%%%,
%
\section{Observations}\label{S:obs}
The original plan of the program was to maximize the overlap with the 
deep optical \textit{HST} data of GO-10146 (PI: Bedin), which allowed us to study 
the entire WD\,CS of M\,4 \citep{2009ApJ...697..965B}. 
This strategy would have provided both proper motions (to assess the cluster membership) 
and the optical counterparts of our \textit{JWST} data. 
However, our GO-1979 \textit{JWST} observations (PI: Bedin) were accepted well before the \textit{JWST} 
launch and, as observations of other GCs became publicly available, our careful examination of \textit{JWST}'s performances  \citep{2022MNRAS.517..484N,2023AN....34430006G,2023MNRAS.525.2585N} revealed that the area covered by 
the GO-10146 \textit{HST} data was too crowded for \textit{JWST} (see end of Sect.\,\ref{subsec:HST}). 
Internal reflections and diffraction spikes make the background level too high to detect faint sources on NIRCam 
images, hampering our ability to measure the faintest stellar objects in that \textit{HST} field.
Therefore, we redesigned the observations shifting the field further from the cluster's centre where crowding 
is less severe, at the cost of a smaller overlap with the \textit{HST} data and, in turn, a smaller sample of stars 
with proper motions (for the cluster membership) and optical counterparts.
%%%%%%%%%%%%%%%%%%%%%%%%%%%%%%%%%%%%%%%%%%%%%%%%%%%%
%
\subsection{\textit{JWST} images}

The GO-1979 images were taken on April 9, 2023 (epoch $\sim$2023.27) using a 6-point \texttt{FULLBOX} primary 
dither pattern with \texttt{2-POINT-LARGE-WITH-NIRISS} sub-pixel positions. Each pointing comprises a single 
image with both short- (SW) and long-wavelength (LW) detectors using a \texttt{MEDIUM8} readout pattern 
with 5 groups/1 integration (effective exposure time of 515.365\,s). 
Thanks to NIRCam's simultaneous observations, we secured images in F150W2 (SW) and F322W2 (LW) filters. 
The top-left panel of Figure\,\ref{fig:FoV} shows a Digital\,Sky\,Survey\,2 InfraRed
\footnote{\texttt{Plate ID: A376} collected 1982/07/28.} 
image of the field around our NIRCam pointing, which is centered 
at $( \alpha; \delta ) = (245^\circ\!.9297;-26^\circ\!.4504)$, 
at an average angular distance from the cluster centre of about 
4.8\,arcmin, ranging between 1.9\,arcmin and 8\,arcmin. 
The top-right panel of Figure\,\ref{fig:FoV} shows the F150W2 
stacked image for the entire NIRCam field of view (FoV), 
while the bottom panel shows a zoom-in of $3^\prime\times0.6^\prime$  for a 
relatively ``sparse'' region, highlighting both the high density of stars and 
high number of background unsolved extra-Galactic sources.

The NIRCam images were processed as described in \citetalias{2024AN....34540039B}, i.e., 
using the official \textit{JWST} calibration pipeline \citep{2023zndo..10022973B} to obtain 
the level-2b \texttt{\_cal} images. We increased the dynamical range of our data by using 
the so-called \texttt{frame\,zero} (i.e., the first frame of each integration), which enables 
the ramp fit in pixels that saturate after the first group. We modified the \texttt{\_cal} images by 
\textit{(i)} converting the values of the pixels from MJy sr$^{-1}$ into counts using the 
necessary header keywords in each \texttt{FITS} file, and 
\textit{(ii)} flagging unusable pixels using the data quality (\texttt{DQ}) flags 
available in each \texttt{\_cal} multi-extension \texttt{FITS} file.

\subsection{\textit{HST} images}
\label{subsec:HST}
The \textit{HST} data are from GO-10146 \citep{2004hst..prop10146B} taken with the 
Wide Field Channel (WFC) of the Advanced Camera for Surveys (ACS). The observations were split into two parts. 
The first part was obtained between 2004 July and August ($\sim$2004.607) in the F606W filter 
(20 exposures of $\sim$1200\,s each). The second part was secured in June 2005 and 
included only four exposures (of $\sim$1200\,s each) in the F775W filter.\\~ 

A small digression here: 
we want to emphasize that the diffraction pattern of the PSFs in \textit{JWST} images 
has profound effects on sky-background-limited detection of sources in crowding conditions, 
such as those of the cores of Galactic GCs. 
To exemplify this, in Fig.\,\ref{fig:Zom} we show a common portion of 
the two \textit{JWST} and  \textit{HST} datasets. 
The left-panel shows the finding chart in the NIRCam field of this selected 
region (indicated by a yellow rectangle); this region is in the most crowded part of the 
\textit{JWST}'s NIRCam FoV, and the least crowded portion of the \textit{HST}'s ACS/WFC FoV. 
The middle panel shows this selected region as observed by \textit{JWST} in NIRCam/SW filter F150W2, 
while the right panel shows the same region as observed in \textit{HST} ACS/WFC filter F775W. 
The scales in the two rightmost panels of this figure have a stretched contrast to enable visibility of the sky-noise levels. 
Although not quantitative, this figure qualitatively shows that even in the darkest region in the \textit{JWST} images, 
there are always diffraction spikes coming from relatively distant bright sources, 
effectively compromising the sky-background level. Indeed, to be detected, faint 
stars will have to compete not only against the local sky-level noise but also against 
light coming from far away sources, \textit{de facto} hampering to detect easily 
sources just above the local sky background.

%%%%%%%%%%%%%%%%%%%%%%%%%%%%%%%%%%%%%%%%%%%%%%%%%%%%
%
\section{Data Reduction}\label{S:DR.msrmt}
%
%%%%%%%%%%%%%%%%%%%%%%%%%%%%%%%%%%%%%%%%%%%%%%%%%%%%
%
\subsection{\textit{JWST}'s astrometry and photometry} 
All images obtained with NIRCam were reduced using software tools and methods described 
in detail in the first three papers of the series \textit{``Photometry and astrometry with \textit{JWST}''} 
\citep{2022MNRAS.517..484N,2023AN....34430006G,2023MNRAS.525.2585N}, and applied with success 
in the study of brown dwarfs of 47\,Tucan\ae\, by \citet{2023MNRAS.521L..39N}, 
and more recently in \citetalias{2024AN....34540039B}. 
The method consists of a \textit{first-} and \textit{second-pass}  photometry as defined by \citet{2008AJ....135.2055A}. 

Briefly, the \textit{first-pass} photometry measures the positions and fluxes of all detectable sources in each NIRCam 
image via effective-PSF fit (see \citetalias{2024AN....34540039B} and \citealt{2023MNRAS.525.2585N} for a description 
of the effective-PSF models). Positions were corrected for geometric distortion using the solution provided 
by \citet{2023AN....34430006G}. These astro-photometric single-image catalogues were then transformed onto a 
common reference frame, which was established using the \textit{Gaia}\,DR3 catalogue  
(after the \textit{Gaia} positions were transformed to the epoch of our \textit{JWST} observations). 
Placing all first-pass outputs onto the same system allows them to be properly 
compared during the next step of the data reduction.

The \textit{second-pass} photometry takes advantage of the information stored in all images simultaneously 
and improves the detectability and the photometry of very faint sources that would otherwise be lost in an individual image. 
In our work, the second-pass photometry was performed with a modified version of the code \texttt{KS2}, 
developed by Jay Anderson and described in \citet[and references therein]{2017ApJ...842....6B}. 
Along with fluxes and positions, \texttt{KS2} produces several quality diagnostics, 
including the root mean square (RMS) error in brightness (in magnitudes), 
the PSF quality of fit (\texttt{q}), and a ``stellarity index'' that describes how well the shape of a 
given source resembles that of the PSF (the so-called \texttt{RADXS}; see \citealt{2008ApJ...678.1279B}).

%__________________________________________________________________
\begin{figure}
\centerline{\includegraphics[width=88mm]{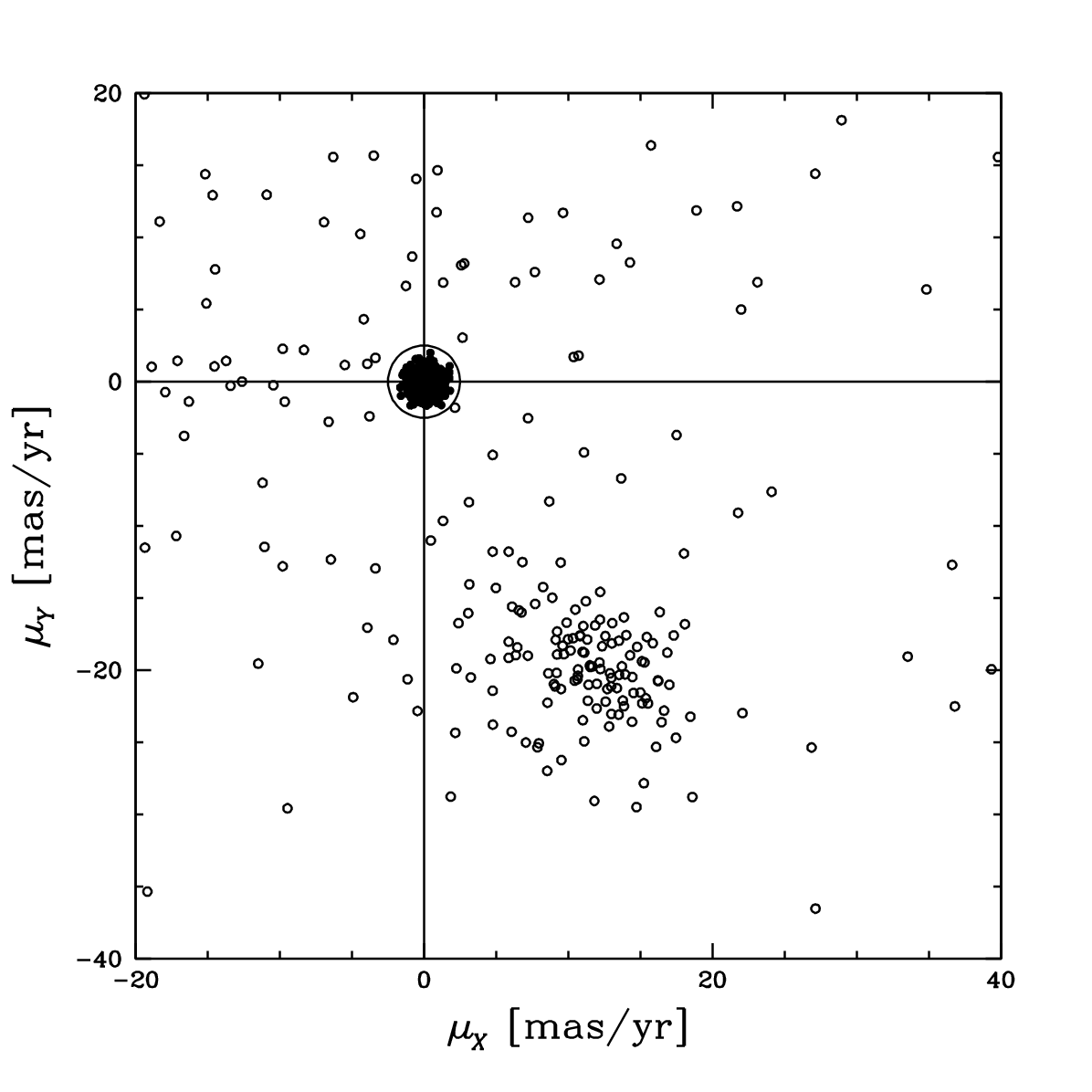}}
\caption{
Vector point diagram for sources that passed all selection criteria based on \texttt{KS2} diagnostics. 
The time baseline is 18.66\, years.
Proper motions (oriented along NIRCam' detectors) and are computed with respect to the cluster members, therefore the coordinates (0;0) 
coincide with the average motion of the stellar systems (indicated by the vertical and horizontal lines for the axes' origin).
The inner circle corresponds to a proper-motion membership criterion for M\,4's stars 
(filled symbols), defined as 2.5\,mas\,yr$^{-1}$ (see text). 
}
%%%
\label{fig:VPD}
\end{figure}
%__________________________________________________________________
%

%%%%%%%%%%%%%%%%%%%%%%%%%%%%%%%%%%%%%%%%%%%%%%%%%%%%%%
\subsection{\textit{HST}'s astrometry and photometry} 
%%%%%%%%%%%%%%%%%%%%%%%%%%%%%%%%%%%%%%%%%%%%%%%%%%%%%%
%
The data reduction for \textit{HST} data was performed by \cite{2009ApJ...697..965B}, with 
methodologies that are the precursor to those employed for \textit{JWST} data and briefly described in that work, 
but are fully detailed in the seminal study by \cite{2008AJ....135.2055A}.

\subsection{Calibrations of astrometry and photometry} 
We calibrated the \textit{JWST} and \textit{HST} photometry to the Vega-magnitude system following the procedures 
outlined by  \cite{2023MNRAS.525.2585N} and \citet[e.g.,][]{2005MNRAS.357.1038B}, respectively. 
Hereafter, we will adopt the following symbols for the calibrated magnitudes: 
$m_{\rm F606W}$, 
$m_{\rm F775W}$, 
$m_{\rm F150W2}$, and
$m_{\rm F322W2}$.  
Positions were registered onto the International Celestial Reference System (ICRS) frame 
using the \textit{Gaia}\,DR3 catalogue   (after the \textit{Gaia} positions were transformed 
to the epoch of the \textit{HST} or \textit{JWST} observations).

\subsection{Proper Motions}  
\label{Sect:pms}

Proper motions were computed as the positional displacements of the sources between 
the \textit{JWST} and \textit{HST} epochs divided by the temporal baseline (approximately 18.66\,years). 
We adopted a NIRCam SW pixel scale of $\sim$31.2\,mas \citep{2023AN....34430006G}.
PMs are relative to the bulk motion of the cluster members in the field, 
meaning that the cluster distribution in the vector-point diagram is centred at the origin.
[The second (much loose) clustering of sources at $\sim20$\,mas\,yr$^{-1}$ are 
mainly stars of the Galactic Bulge \citep{2003AJ....126..247B}]. 
These PMs are oriented with the instrumental NIRCam coordinates. 

In the studied field of M\,4, stars have an internal dispersion of at 
most $\sim$5\,km\,s$^{-1}$ \citep{2021MNRAS.505.5978V,2022ApJ...934..150L,2023MNRAS.522.5740V}, 
which at a GC distance of $\sim$1850\,pc \citep{2021MNRAS.505.5957B} corresponds to a PM dispersion 
of less than 0.5\,mas yr$^{-1}$, i.e., a displacement lower than 10\,mas $\simeq$0.3 NIRCam SW pixel 
over 18.66\,yr. This value is considerably smaller than the random PM uncertainties at the 
faint magnitudes of the bulk of the WDs in our sample;  
estimated to be about 1\,NIRCam/SW pixel per 
epoch (2.5\,mas/yr$\sim$1\,pxl$\times\sqrt{2}$\,epochs$\times$31.2\,mas/pxl/18.66\,yr).  
Hereafter, we safely consider as cluster members all sources with a PM 
lower than 2.5\,mas yr$^{-1}$ from the bulk motion of the cluster 
in the vector-point diagram (see Fig.\,\ref{fig:VPD}).

%%%%%%%%%%%%%%%%%%%%%%%%%%%%%%%%%%%%%%%%%%
\subsection{Artificial Stars} 
\label{Sect:ASTs}
%%%%%%%%%%%%%%%%%%%%%%%%%%%%%%%%%%%%%%%%%%

Artificial-star tests (ASTs) are an important step in assessing the reliability of 
point-source photometry and the completeness of our sample.
We generated 100,000 artificial stars homogeneously distributed within the NIRCam FoV 
and with a uniform F150W2 magnitude distribution in the magnitude range of the WD CS. 
We then used the \texttt{KS2} software to find and measure these artificial stars 
in the \textit{JWST}/NIRCam images just as we did for real stars. 
The comparison between the input and output parameters (positions and magnitudes) provides us with 
an estimate of the reliability of our photometry.

%__________________________________________________________________
\begin{figure*}
\centerline{\includegraphics[width=168mm]{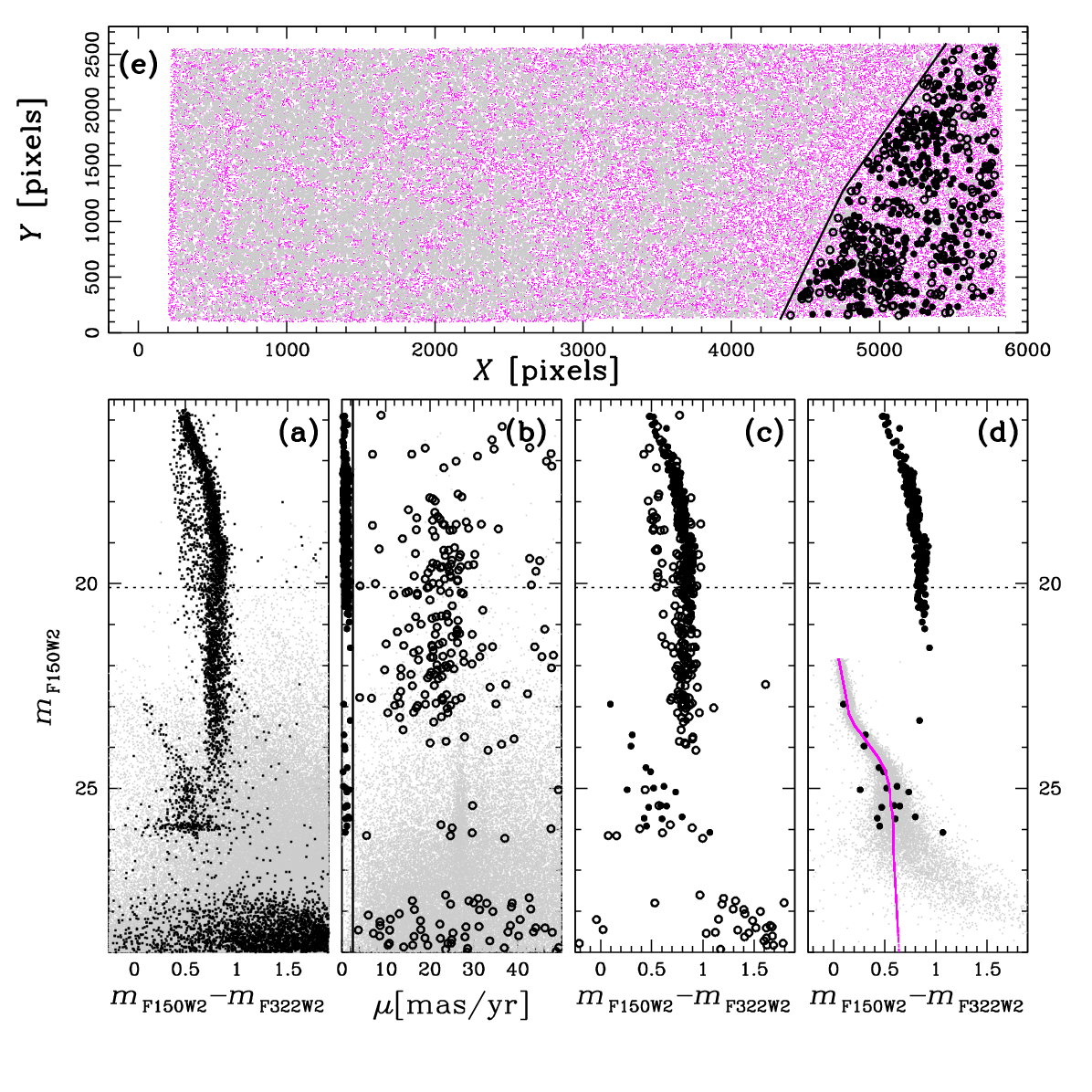}}
\caption{--- 
\textit{(a)}:  CMD of all sources detected within NIRCam (grey dots) and of those passing the quality test (black dots). 
The dashed horizontal lines in all bottom panels indicate the onset of saturation for objects brighter than $m_{\rm F150W2}\simeq20$.
\textit{(b):} For all sources with two epochs the total-PMs vs.\,$m_{\rm F150W2}$ (grey dots). 
Those passing quality selections are indicated in black symbols. 
A vertical line defines our membership criterion defined in Sect.\,\ref{Sect:pms} (2.5\,mas\,yr$^{-1}$). 
Those with proper motions below that line are considered members (filled circles), and those with larger motions  
are considered field objects (open circles), and their CMD is shown in panel \textit{(c)}. 
\textit{(d):} The CMD for members (filled circles) and artificial stars, these as added (in magenta) and as recovered (in grey).
\textit{(e):} The spatial distribution of sources in \textit{(c)} and \textit{(d)} across the NIRCam FoV.
The black line marks the region of overlap with ACS/WFC observations from GO-10146 overlaps, 
available on the right of that line (see, Fig.\,\ref{fig:FoV}).  
}
%%%
\label{fig:JWST-CMD}
\end{figure*}
%__________________________________________________________________
%

\begin{figure*}
\centerline{\includegraphics[width=15truecm]{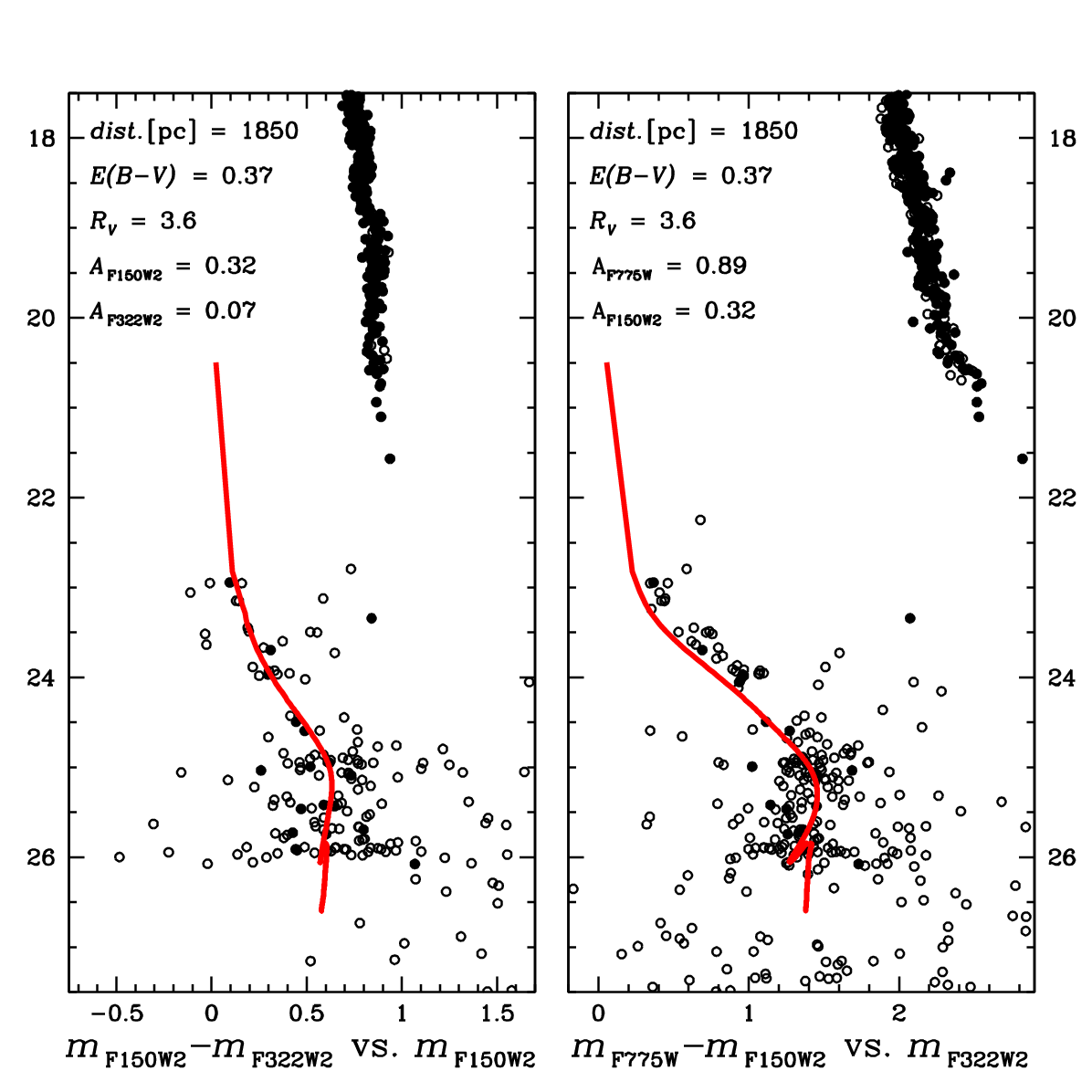}}
\caption{
The two most informative CMDs, for the WD CS in M\,4, 
F150W2$-$F322W2 vs.\,F150W2 (left), and F775W$-$F150W2 vs.\,F322W2 (right). 
In these CMDs we display sources with a PM estimate, and in particular only those with 
PMs within 1.7\,mas/yr the cluster's bulk motion (open-circles).  
Those that \textit{also} passed the quality-check parameters are indicated by filled circles. 
%	data "models/IWDnewBaSTIZ004aent12000_c07opa.DA.jwst" read {F150W2  8 F322W2 12}
%	data "models/IWDnewBaSTIZ004aent12000_c07opa.DA.acs"  read { F606W 10 F775W  14}
The red lines show the 12-Gyr DA WD isochrone from \cite{bastiwd}. 
}
%%%
\label{fig:bestCMD}
\end{figure*}
%__________________________________________________________________
%

%%%%%%%%%%%%%%%%%%%%%%%%%%%%%%%%%%%%%%%%%%%%%%%%%%%%
%
\section{The \textit{JWST} CMD of M\,4}
\label{S:CMDs}
%
%%%%%%%%%%%%%%%%%%%%%%%%%%%%%%%%%%%%%%%%%%%%%%%%%%%%

%
By using only 14 \textit{HST} orbits, \cite{2009ApJ...697..965B} 
were able to observe the entire CMD from the red-giant branch (RGB) tip to the location of 
the peak of the WD luminosity function (LF) at the bottom of the CS in M4, located at $m_{\rm F606W}$=28.5$\pm$0.1.
The CMD from the MS to the horizontal branch stage was employed to determine the distance modulus, extinction, 
and MS TO age from isochrone fitting.  
A comparison of the observed white dwarf luminosity function
with results from theoretical models provided an independent age determination.
The age inferred from the white dwarf luminosity function 
was 11.6$\pm$0.6\,Gyr, in good (internal) agreement with the age from fitting 
the MS\,turn off (TO) (12.0$\pm$1.4\,Gyr). 
The error bars took into account only the intrinsic errors of the methods and observations, 
not uncertainties in the adopted theoretical models.\\~  
Here, for the first time, 
thanks to \textit{JWST}, 
we have been able to extend the observations of the entire WD\,CS to the IR; 
where reddening variations are less significant for a GC with a complicated 
foreground field extinction such as M\,4 \citep{2012AJ....144...25H}.\\~ 
%%%% 
%%% https://iopscience.iop.org/article/10.1088/0004-6256/144/1/25
%%%%
 
In panel \textit{(a)} of Fig.\,\ref{fig:JWST-CMD}, we present the CMD $(m_{\rm F150W2} - m_{\rm F322W2})$ \textit{vs.\ } $m_{\rm F150W2}$  
for all the local maxima detected across the entire NIRCam FoV.
All detected sources are shown in grey, while those that were well measured and passed all selection 
criteria for the  \texttt{KS2} quality parameters (cfr., \citetalias{2024AN....34540039B}) are marked in black. 
The CMD extends over ten magnitudes in brightness thanks to the NIRCam \texttt{frame\,zero}; 
however, sources with magnitudes above the saturation level (marked by the horizontal dashed line at about $m_{\rm F150W2} \simeq 20$) 
should be regarded with caution, as probably affected by systematic photometric errors (not yet quantified). 
Panel \textit{(b)} displays the \textit{total} PMs (i.e., the sum in quadrature of the PMs in the two 
components $\mu_X$ and $\mu_Y$) vs.\  $m_{\rm F150W}$ for a subset of the source in \textit{(a)} 
(those detected in both \textit{JWST} and \textit{HST}). 
All sources are shown as grey dots, while those that passed the quality parameters are indicated by black symbols.
Among these, we define as cluster members those sources that meet the selection criterion defined in 
Sect.\,\ref{Sect:pms} (i.e., PMs$<$2.5\,mas\,yr$^{-1}$, and indicated by a vertical line), 
and show them with filled circles, while we use open circles for likely field objects. 
The CMD of these sources is shown employing the same symbols in panel \textit{(c)}. 
Panel \textit{(d)} shows only the member stars (filled circles) and the artificial stars 
(as added in magenta, and as found in grey) passing the same quality-parameter selections passed by real stars.  
Finally, the top panel of Fig.\,\ref{fig:JWST-CMD} illustrates the spatial distribution 
of objects from panel \textit{(c)} and \textit{(d)}, employing the same symbols.  
At the bottom-right of this panel, a black line marks  the boundary of the region in common between \textit{JWST} and \textit{HST} (to the right of this line).\\~

Although, the WD\,CS of M\,4 is partially lost in the midst of the field objects, 
its location can be clearly identified in the CMD presented in Panel \textit{(a)} of Figure\,\ref{fig:JWST-CMD}, 
a location that is confirmed by those few stars with PMs membership [filled dots in panels (d) and (e)].
Even though the lack of PM memberships for all sources within the entire NIRCam FoV,  
prevents a detailed quantitative study of the exact shape of the WD\,CS luminosity function of M\,4, 
a clear feature emerges. 
This is the over density at magnitude 
$m_{\rm F150W2} \sim 26$ and  color $(m_{\rm F150W2}-m_{\rm F322W2})\sim0.5$, which 
can not be reproduced by any other population of the Galactic field or background sources; 
it must correspond to the peak of the WD\,CS luminosity function. 
This feature is well above the completeness drops, and it is a feature 
that models can predict with great accuracy; 
indeed it aligns with the location predicted by the models.
Also, note that essentially all background galaxies were removed by the selection in stellarity parameters (\texttt{RADXS}, i.e., 
how well the sources profiles resemble the PSFs).  
Therefore, a histogram distribution in magnitude (with a 0.1\,mag step) for stars along the WD\,CS of M\,4, provides 
an accurate estimate of the  luminosity of the white dwarf luminosity function peak, which is firmly located at magnitude $m_{\rm F150W2}=25.90\pm0.10$
(and comforted by visual inspection of panel \textit{(a)} and \textit{(d)} of Figure\,\ref{fig:JWST-CMD}).\\~

A major difference between optical and IR CMDs of a 
GC CS is that in the optical the CS  
displays a characteristic `blue turn' \citep[e.g.,][]{2013A&A...549A.102B}, 
which corresponds to the appearance of increasingly more massive and thus smaller radius WDs originating from shorter-lived progenitors. 
This blue turn, depending on the cluster age --hence of the $T_{\rm eff}$ of the coolest WDs-- may be further enhanced by the onset of 
collision-induced absorption (CIA) of H$_2$ in the atmospheres of the coolest WDs \citep{1998MNRAS.294..557H}.  
In the optical-IR CMD, obtained with the \textit{JWST}/NIRCam filters F150W2 and F322W2, 
combined with the optical \textit{HST}/ACS/WFC/F775W filter, the effect of CIA is evident and unambiguous. 
It begins at higher $T_{\rm eff}$ compared to 
optical CMDs, when the isochrone still aligns with the cooling track of single low-mass ($\sim 0.55M_{\mathSun}$) WDs.

This transpires clearly from Fig.\,\ref{fig:bestCMD}, 
which shows an IR and a mixed optical-IR CMD of 
the cluster CS, together with a representative theoretical 
12\,Gyr WD (with hydrogen atmospheres) isochrone from 
progenitors with [Fe/H]=$-$1.0 \citep[from][]{bastiwd}.
The isochrone has been shifted by the labelled 
distance \citep[from][]{2021MNRAS.505.5957B} and mean 
reddening \citep[from][]{2012AJ....144...25H}, employing $A_{\lambda}/A_{V}$ ratios calculated 
as in \citet{2005MNRAS.357.1038B}\footnote{We used the extinction law by 
\citet{gordon} that covers the wavelength 
range of our chosen IR filters.} using 
$R_V$=3.6 \citep[from][]{2012AJ....144...25H}.

Whilst the IR CMD (and the isochrone) shows a sequence 
that at the faint magnitudes is almost vertical, 
the optical/IR CMD displays a turn to the blue 
corresponding to the portion of the isochrone populated by constant mass ($\sim 0.55M_{\mathSun}$) objects, 
hence caused by the onset of CIA, as we have seen in NGC\,6397.

%__________________________________________________________________
\begin{figure}
\centerline{\includegraphics[width=8.9cm]{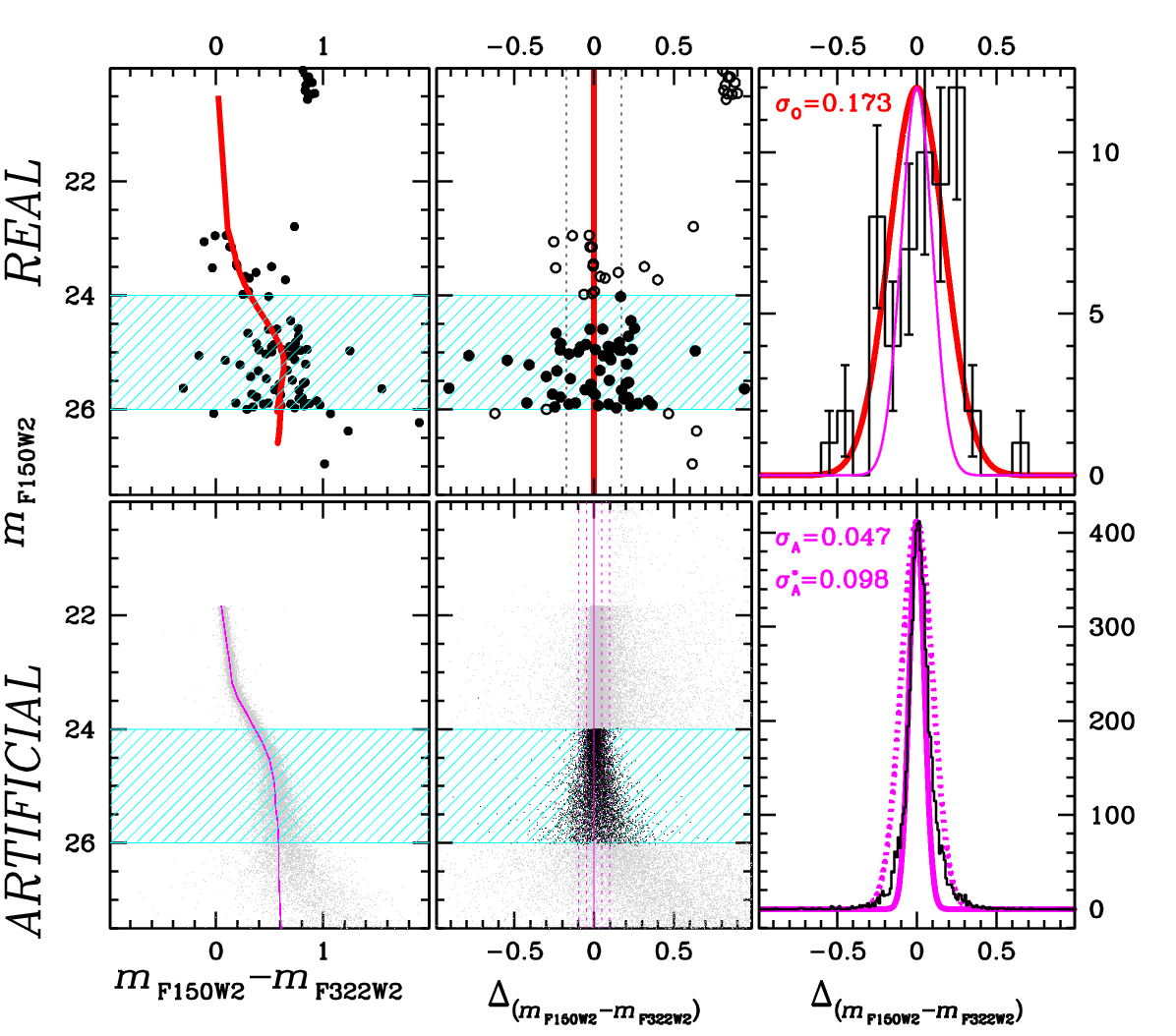}}
\caption{
\textit{(Top):} Real stars.  
\textit{(Bottom):} Artificial stars.  
\textit{(Top-Left):} CMD for all sources with PM within 1\,mas\,yr$^{-1}$ of the cluster. 
The red line is the same isochrone used in Fig.\,\ref{fig:bestCMD}. 
\textit{(Top-Middle):} The rectified CMD along the isochrone. The histogram for the 
sources within the magnitude interval indicated by the shaded area in cyan (filled dots), is shown 
on the \textit{(Top-Right)} panel. The dispersion value of these observed data points ($\sigma_{\rm O}$) 
was fitted with a Gaussian, whose $\sigma$ is indicated in red. 
\textit{Bottom} panels show the same for artificial stars. 
The magenta line is the line along which the ASTs were added. 
We do not have PMs for ASTs, therefore we show the two dispersions values (indicated in magenta) obtained 
for all ASTs ($\sigma^*_A$) and for those ASTs that have passed all the photometric quality tests ($\sigma_A$).
}
%%%
\label{fig:IRexcess}
\end{figure}
%__________________________________________________________________
%

%__________________________________________________________________
\begin{figure*}
\centerline{\includegraphics[width=\textwidth]{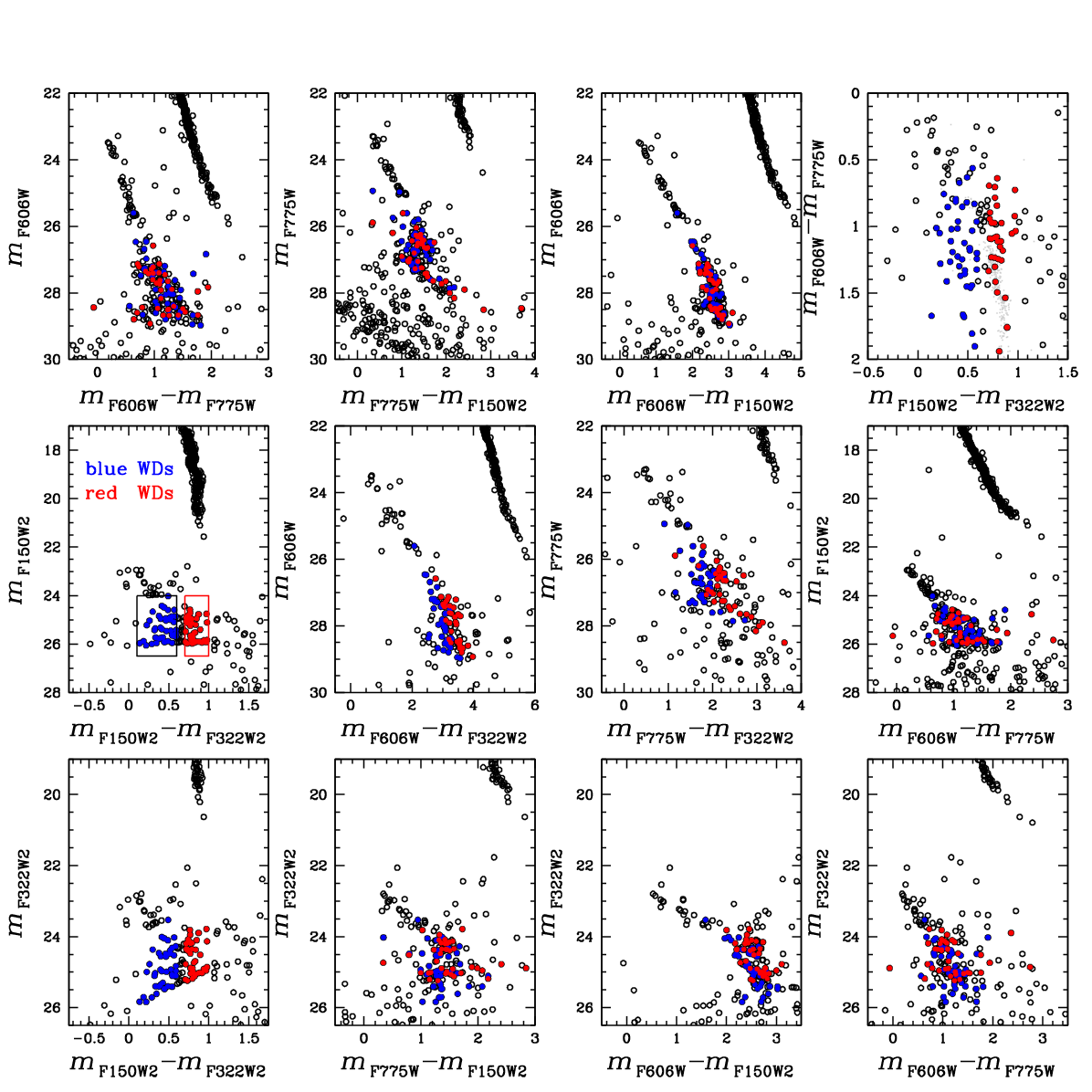}}
\caption{
Eleven CMD combinations and one TCD from the four \textit{HST} and the \textit{JWST} filters employed in this study.  
These figures highlight the red and blue WD samples defined in the F150W2$-$F322W2 
versus F322W2 CMD (leftmost panel in the middle row). In the TCD we have marked as 
grey-small dots the main sequence stars (top-right panel). 
In summary, the colour dispersion of the faint end of the WD\,CS of M\,4 appears considerably larger than what 
expected on the basis of the photometric errors, but this excess seems to be present only in filter F322W2. 
Those sources, defined as red WDs, appear to show an excess only in filter F322W2 (by as much as $\sim$0.5\,mag).  
}
%%%
\label{fig:BRcmds}
\end{figure*}
%__________________________________________________________________
%

%%%%
%
\subsection{Possible red excess WDs in M\,4?} 
\label{sect:IRexc}

Given the relatively straightforward modeling of cool WD photospheres in the infrared bands, 
we use these NIRCam observations to search for infrared excess that could be caused 
by circumstellar debris disks \citep{2011AJ....142...75C}. 
Disk-induced infrared excess is well documented among young WDs in the Milky Way \citep{2007ApJ...657L..41S}, 
and finding debris disks around WDs in globular clusters would suggest favourable conditions for planet 
formation in these ancient systems, thus making such a discovery especially significant. 
Recently, an infrared excess was hinted at for the first time in a GC in \textit{JWST} observations of 
WDs in NGC\,6397 \citepalias{2024AN....34540039B}. 
While this feature requires further confirmation ---being significant only in the F322W2 filter and 
more pronounced at fainter magnitudes--- it would be valuable to investigate whether similar features 
are also present in a second cluster: M\,4.

However, the \textit{JWST} data for M\,4 are of substantially lower quality than those used for NGC\,6397 
in this investigation for two main reasons. First, membership could only be established in a much smaller region, 
and second, this region lies too close to the cluster core, where crowding significantly increases photometric errors.\\~

In Fig.\,\ref{fig:IRexcess} we intend to demonstrate first that the 
dispersion along the observed WD\,CS of M\,4 is significantly larger 
than what is inferred by our estimated uncertainties. 
The top panels refer to real sources, while the bottom ones to artificial stars. 
In the left panel we show the CMDs, in the middle ones the rectified CMD along the 
reference WD\,CS, and on the right ones the histogram distributions in magnitude
for stars in a very specific magnitude interval, i.e., between $m_{\rm F150W2}=$24--26 (area shaded in cyan). 
In these histograms we show the Gaussians obtained taking as dispersion the 68.27$^{\rm th}$ percentile of the distributions. 
In red the Gaussian for the observed sources (with labeled in the same colour the corresponding values of $\sigma_{\rm O} = 0.173$\,mag), 
while in magenta the Gaussian for the artificial stars. 
Not having reliable PMs estimates for the ASTs, we estimated the dispersion for the entire sample ($\sigma^*_{\rm A} = 0.098$\,mag) 
and for those ASTs passing the all the quality selections ($\sigma_{\rm A} = 0.047$\,mag). 
The observed dispersion is significantly larger than the estimated one even in the worst case ($\sigma_{\rm O} > \sigma_{\rm A})$.
Therefore, assuming that errors are well understood, we note that the observed dispersion of the final portion of the  WD\,CS of M\,4 
has a significantly broader color than expected on the basis of photometric errors (by more than 50\%).

Furthermore, assuming that the isochrone is `the truth', 
we can quantify the number of WDs on the red side and on the blue side of this 
isochrone within 3$\sigma_{\rm O}$. For stars with proper-motions within 2.5\,mas\,yr$^{-1}$ from
the cluster members, we count 63 red WDs and 38 blue WDs, 
providing a $\sim$2.5\,$\sigma$ significance for the excess-number of red-WDs. 
Limiting the sample to stars with proper-motions within 1\,mas\,yr$^{-1}$, 
significance becomes $\sim$2\,$\sigma$.
\\~   

Next,  analogous to the approach in \citetalias{2024AN....34540039B}, we \textit{arbitrarily} define two regions on the CMD 
$(m_{\rm F150W2}-m_{\rm F322W2})$ vs.\,$m_{\rm F150W2}$: one on the red side and one on the blue side of the isochrone, 
around the nearly vertical section between $m_{\rm F150W2}=24$ and 26.5. 
We then investigate whether the two samples of blue and red WDs, 
defined by the two regions indicated in the mid panel on the left in Fig.\,\ref{fig:BRcmds}, 
would correspond to distinct locations in other CMDs generated by combining all available filters in the 
overlapping \textit{HST} and \textit{JWST} FoVs. 
The results of these tests are shown in Fig.\,\ref{fig:BRcmds},  
displaying eleven CMDs and one two-colour diagram (TCD). 
% R: acryonim used in the caption.
% 
As observed for NGC\,6397, the two samples can be disentangles only when the F322W2 filter is involved. 
This suggests that any potential IR excess --as large as $\sim$0.5\,magnitudes-- among WDs appears solely in the F322W2 filter.

In summary, the $(m_{\rm F150W2}-m_{\rm F322W2})$-colour dispersion of the faint end of the WD\,CS of M\,4 appears considerably larger than what 
expected on the basis of the photometric errors (by as much as 50\%; likely more). However, this excess seems to be present only in filter F322W2. 
Furthermore, while in the case of NGC\,6397 \citepalias{2024AN....34540039B} there was an indication of a split of the faint part of the WD\,CS, in the case of M\,4, 
there is not even a hint for such a feature. The lack of a similar indication could be 
due to the considerably lower quality of the M\,4 data set 
(membership available only for sources in a small field of high crowding).

Finally, although we consider it unlikely, 
we can not confidently exclude the presence of yet unidentified large errors in filter F322W2 that our artificial star tests can not track. Such error could potentially inflate the observed dispersion of the faintest part of the WD\,CS.\\ 

Follow-up observations are needed to confirm this 
broader-colour distribution than expected for the faint end of the WD\,CS of M\,4,  
for members across the whole NIRCam field of view (particulary in less crowded regions), and
at even redder wavelengths to better understand the nature of these IR excesses. 
At a minimum, a second epoch with observations in the F444W filter of NIRCam 
are required to verify this feature, ideally complemented by MIRI low-resolution spectra.

%%%%%%%%%%%%%%%%%%%%%%%%%%%%%%%%%%%%%%%%%%%%%%%%%%%%
\subsection{Comparison between M\,4 and NGC\,6397} 
\label{sect:CIA}
%%%%%%%%%%%%%%%%%%%%%%%%%%%%%%%%%%%%%%%%%%%%%%%%%%%%%

As already mentioned, we found that in NGC\,6397 the peak of the luminosity function 
at the bottom end of the CS is located at $m_{\rm F150W2}=26.55\pm0.10$, whilst in the case of 
M\,4 we have estimated $m_{\rm F150W2}=25.90\pm0.10$, corresponding to a difference $\Delta m_{\rm F150W2}=0.65\pm0.14$\,mag.

If we take into account the different distances (that have a negligible formal error compared to 
the error on the observed difference of the peak magnitudes) and extinctions [$E(B-V)$ variations up 
to $\pm$0.05-0.06\,mag around the adopted mean values for the two clusters provide a 
negligible contribution to the error budget on the luminosity function peak magnitudes], 
the peak of M\,4 WD\, luminosity function is 0.24$\pm$0.14\,mag brighter than the NGC\,6397 counterpart. This corresponds 
to an age difference $\Delta t$ equal to 0.8$\pm$0.5\,Gyr, using \citet{bastiwd} WD models.
Considering the error on $\Delta t$ we find that M\,4 is only very marginally younger than NGC\,6397, if at all.

Such a small value of $\Delta t$ is broadly consistent with the differences between the clusters'  main sequence turn-off ages obtained by  \citet{sw} and \citet{mf}, who found the two clusters virtually coeval within the errors (errors equal to 1.1\,Gyr for \citealt{sw}, and 0.7\,Gyr for \citealt{mf}), whilst 
the turn-off ages determined by \citet{vdb} give a more significant age difference $\Delta t$=1.5$\pm$0.5, M\,4 being younger.

In terms of absolute age our derived value of $\Delta t$ translates to 12.2$\pm$0.5\,Gyr for M\,4, 
consistent with the best estimate of 12.1\,Gyr derived by \citet{hans04M4} from their analysis of M\,4 white dwarf cooling sequence.

The absolute turn-off based ages of M\,4 derived by 
\citet{sw} and \citet{mf} are in the range 11.7--11.9\,Gyr and 12.7--13.3\,Gyr (with the previously quoted errors) respectively, the exact values depending on 
adopted [Fe/H] scale and set of theoretical isochrones, 
while \citet{vdb} obtain 11.5$\pm$0.4\,Gyr.
%
%%% https://people.smp.uq.edu.au/HolgerBaumgardt/globular/fits/ngc6121.html
%%% https://people.smp.uq.edu.au/HolgerBaumgardt/globular/fits/ngc6397.html

%%%%%%%%%%%%%%%%%%%%%%%%%%%%%%%%%%%%%%%
%
\section{Conclusions}

This study offers our first in-depth view of the M\,4 system at infrared wavelengths. 
Future studies will build on these data to explore the main sequence (MS) population 
into the substellar regime, analyze field populations surrounding the cluster, and examine cluster kinematics. 
In this work, we focused on the IR counterpart of the complete white dwarf cooling sequence in M\,4, previously observed 
in the optical through deep \textit{HST} observations \citep{2009ApJ...697..965B}. 
The results of this work can be summarized as follows:
\begin{itemize}
    \item 
    To verify the current predictions of CIA effects along the WD\,CS for only the second time in a 
    globular cluster (after NGC\,6397, \citetalias{2024AN....34540039B}), we focus on M\,4, a cluster with 
    significantly different chemical composition, age, and kinematic status compared to NGC\,6397. 
    Our observations shows that the model predictions are consistent with the data within the uncertainties, 
    also for M\,4.
    \item 
    Searching for WDs with possible IR excess, we found that the WDs with magnitudes between  $m_{F322W2}=24$ and 26
    exhibit a colour distribution significantly broader ($\sim$0.17\,mag) than the expected 
    errors ($\sim$0.10\,mag), and 
    somehow compatible with the recent findings in \citetalias{2024AN....34540039B} for NGC\,6397, where approximately 25\% displayed IR excess. 
    In the case of M\,4, however, while the breadth of the bottom part of the WD\,CS 
    is also a notable and significant feature ---as 50\% larger 
    than the expected errors---  there is no indication for a separation between two populations.  
    This could be due to the limited sample of members
    and to the degraded photometry in the dense cluster core region probed, 
    the only area for which proper motion memberships were available. 
    \item 
     By comparing the absolute F150W2 magnitudes 
     of the luminosity function peak at the bottom of the observed  WD\,CSs in M\,4 and NGC\,6397, 
     we find only marginal evidence suggesting that M\,4 is slightly younger, by 0.8$\pm$0.5\,Gyr. 
     This age difference translates to an absolute age 12.2$\pm$0.5~Gyr for M\,4.
    \item 
    Finally, one of our goals is to make publicly available our reduced data, high-accuracy photometric catalogue 
    for members, and atlases in multiple filters as part of this article’s supplementary online material.\footnote{
        % open href
        \href{https://web.oapd.inaf.it/bedin/files/PAPERs_eMATERIALs/JWST/GO-1979/P05/}{
               \texttt{https://web.oapd.inaf.it/bedin/files/PAPERs\_eMATERIALs/\-JWST/GO-1979/P05/}
        % close href
             }
        % close footnote
                             } 
    Further theoretical investigations by other independent groups might find these data useful.
\end{itemize}

\noindent
We plan to extend these observations to additional long-wavelength NIRCam filters (e.g., F444W) and 
possibly to MIRI at 10\,$\mu$m (F1000W). This will enable us to confirm and eventually spectroscopically 
characterize the infrared excess detected around a significant fraction of M4 WDs. 
This approach could help disentangle the possible contributions from sub-stellar companions and/or
dust emission \citep{2005ApJ...635L.161R,2023MNRAS.526.3815S}.
%(Reach et al.\ 2005, Swan et al.\, 2023). 
%https://www.spitzer.caltech.edu/image/ssc2006-04a-evidence-for-comets-found-in-dead-star
%https://jwst-docs.stsci.edu/jwst-mid-infrared-instrument/miri-instrumentation/miri-filters-and-dispersers
%https://jwst-docs.stsci.edu/files/97977594/182258950/1/1669853670648/miri_img_pces.png
%https://ui.adsabs.harvard.edu/abs/2005ApJ...635L.161R/abstract
%
A future \textit{JWST} epoch also will allow us to extend proper-motion membership to 
a larger sample of WD members and to fainter objects, well into the BD sequence.

%%%%%%%%%%%%%%%%%%%%%%%%%%%%%%%%%%%%%%%

%
\section*{Acknowledgments}
\noindent
We warmly thank STScI, our Program Coordinator and Instruments Reviewers 
--Shelly Meyett, Mario Gennaro, Paul Goudfrooij and David Golimowski-- 
for their great support during the review of our problematic observations.
LRB, DN, MG and MSc acknowledge support by INAF under the WFAP project, f.o.:1.05.23.05.05. 
MS acknowledges support from The Science and Technology Facilities
Council Consolidated Grant ST/V00087X/1.
ABu, DA, JA, RG, and ABe, acknowledge support from STScI funding associated with GO-1979. 
We thank an anonymous Referee for the prompt and careful review of our manuscript,  
and for the useful suggestions.
%%%%%%%%%%%%%%%%%%%%%%%%%%%%%%%%%%%%%%%%%%%%%%%%%%%%%%%%%%%%%%%%%%%%%%%%%%%%%%%
%%%%%%%%%%%%%%%%%%%%%%%%%%%%%%%%%%%%%%%%%%%%%%%%%%%%%%%%%%%%%%%%%%%%%%%%%%%%%%%
%%%%%%%%%%%%%%%%%%%%%%%%%%%%%%%%%%%%%%%%%%%%%%%%%%%%%%%%%%%%%%%%%%%%%%%%%%%%%%%

\bibliography{biblio}

\end{document}